\documentstyle[epsfig]{article}
\newcommand{\s}{\mathrm}
\newcommand{\idn}{\hspace{\parindent}}
\newcommand{\be}{\begin{equation}}
\newcommand{\ee}{\end{equation}}
\newcommand{\ba}{\begin{eqnarray}}
\newcommand{\ea}{\end{eqnarray}}
\newcommand{\bb}{\begin{thebibliography}}
\newcommand{\eb}{\end{thebibliography}}

\begin{document}
\title{A Role of the Axial Vector Mesons on the Photon Production in Heavy Ion Collisions and Their Relevant Decays}
\author{Yun Chang Shin$^{1)}$, Myung Ki Cheoun$^{2)}$
\thanks{Corresponding Author, cheoun@phya.snu.ac.kr}, K. S.
Kim$^{3)}$, and T. K. Choi$^{4)}$
\\
1)\it{Cyclotron Application Laboratory, Korea Cancer Center
Hospital(KCCH),}
\\
\it{Seoul,139-706,Korea}
\\
2)\it{IUCNSF, Building 139-1, Seoul National University, Seoul,
151-742, Korea}
\\
3)\it{Department of Physics, Sungkunkwan University, Suwon, Korea}
\\
4)\it{Department of Physics, Yonsei University, Seoul, 120-749
Korea} }
\date{6 August, 2001}
\maketitle
\begin{abstract}
A role of the axial vector mesons, such as $K_1$ and $a_1$, on the
emitted photon spectrum in hot hadronic matter is studied through
the channels $\pi \rho \rightarrow a_1 \rightarrow \pi \gamma$ and
$K \rho \rightarrow K_1 \rightarrow K \gamma$. They are shown to
be dominant channels in this spectrum. This study is carried out
with an effective chiral lagrangian which includes vector and
axial-vector mesons and explains well their relevant decays
simultaneously.\\

\end{abstract}

\section{Introduction}
\idn Recently, the study of dense and hot hadronic matter (HHM) is
one of the interesting fields in relativistic heavy ion physics.
One of the main goals of this study is to find a trace of phase
transition between the hadronic and the quark-gluon plasma (QGP)
phases. Photons (or dileptons as massive photons) can be used as a
reasonable probe to detect this phenomenon because their mean free
paths are much larger than the transverse size of the hot and
dense region of the phases. It means that photons escape without
rescattering, namely, they retain the physical information in each
phase.

Therefore the observation of the QGP signal through the detection
of the photons (or dileptons) could be a plausible choice although
there remained many discussions \cite{ Br92, Ha94, So93, ko96} to
be solved yet, in specific, regarding if the emitted photons are
produced from the HHM or QGP phases.

Yet, to our current knowledge, the HHM and QGP produce energetic
photons and massive photons nearly equally at fixed temperature,
200 MeV, which is a critical temperature for deconfinement and
chiral symmetric phase. Therefore, if any important contributions
have been ignored in the emitted photon spectrum, for example,
strange particles or heavy mesons, it would be very desirable to
investigate those contributions.

Regarding where the photons are emitted from, we consider the
photons produced only through the mesons, i.e, baryon free
matter\cite{Ka91} because baryons are presumed too heavy to be
pair produced at the temperature considered in this report. In
specific, our interest is aimed to find a role of the axial vector
mesons, which usually participated as intermediate states in the
emitted photon spectrum by the mesons.

Actually, since the beginning of '90 \cite{Br92, Ka91,Ka92}, not
only $\rho$ and $\pi$ but also $a_{1}$ meson's role is emphasized
on the channels via mesons, such as $\rho\pi \rightarrow
\pi\gamma$ and $\pi\pi \rightarrow \rho\gamma$ . It should be
noted that $\rho$ and $a_{1}$ form a parity doublet, i.e., they
are chiral partner just like $\pi$ and $\sigma$. But their masses
are not degenerated because of the spontaneous symmetry
breaking(SSB) of chiral symmetry ($\chi$SB). In the chiral
symmetric phase, therefore, $a_{1}$ becomes to be as important as
$\rho$ meson. For instance, Song \cite{So93} and Ko \cite{ko96}
have shown that $a_1$ meson's contribution to $\pi\rho \rightarrow
\pi\gamma$ and $\pi\pi \rightarrow \rho\gamma$ channels could be
dominant within the Massive Yang Mill (MYM) and the gauged linear
sigma models, respectively.

But the $K_1$ ($K \rho \rightarrow K_1 \rightarrow K \gamma$)
meson which could become to be important as the temperature
increases, was not taken into account in their papers. Originally,
the contribution of the $K_1$ meson to the radiative decay in a
hadron gas was studied by Haglin \cite{Ha94}, but with a simple
phenomenological lagrangian introduced by Xiong $et. al$
\cite{Br92}. Moreover their decay widths are overestimated when
they are compared to the experimental data as will be discussed.

In this paper, $K_1$ as well as $a_1$ axial vector mesons'
contributions to the photon spectrum in HHM are calculated and
shown to be dominant channels in this spectrum. This study is
carried out with an effective chiral lagrangian which includes
systematically vector and axial-vector mesons. These fields are
introduced by a non-linear realization of chiral symmetry. This
scheme reproduces well in a consistent manner axial vector mesons'
decays, such as $a_{1} \rightarrow \rho \pi $, $a_{1} \rightarrow
\pi \gamma $, $K_{1}\rightarrow K \rho $ and~$K_{1}\rightarrow K
\gamma $. This successful description was resulted from our
systematic extension of the $SU(2)$ group representation to that
of $SU(3)$\cite{Sh00}.

This paper is organized as follows. In the section 2, our previous
lagrangian \cite{Sh00} is briefly reviewed. The relevant decays of
the axial vector mesons are calculated in section 3 in the
framework of this lagrangian and compared to other calculations
and experimental data available until now. The photon spectrum
through the channels, $\pi \rho \rightarrow a_1 \rightarrow \pi
\gamma$ and $K \rho \rightarrow K_1 \rightarrow K \gamma$ are
investigated in section 4. Brief summary is done at the section 5.

\section{Lagrangian}

\idn Our lagrangian\cite{Sh00} consists of a pseudoscalar meson
sector ${\cal L} ( \pi )$, a spin-1 vector and axial vector meson
sector ${\cal L} ( V, A)$, and a term of interactions with scalar
particles ${\cal L}_{S}$, which comes from mass splitting in the
SU(3) extension of previous SU(2) lagrangian \cite{Pa96}, i.e.,
\begin{equation}
{\cal L} = {\cal L}(\pi ) + {\cal L}(V, A) + {\cal L}_S~.
\end{equation}
The lagrangian for the pseudoscalar meson sector, which is a
leading Lagrangian of the chiral perturbation theory (ChPT), is
given as
\begin{eqnarray}
\cal L(\pi) & = & \frac{f^2}{4} \langle D^\mu U^\dagger  D_\mu U
\rangle +  \frac{f^2 }{4} \langle U^\dagger \chi + \chi^\dagger U
\rangle~,
\\ D_\mu U & = & \partial_\mu U - \it i( \it v_\mu +
\it a_\mu ) U + \it i U ( \it v_\mu - \it a_\mu )~,
\end{eqnarray}
where bracket denotes a trace in flavor space, $f$ is a
pseudoscalar meson decay constant, chiral field is denoted as $U =
\exp( i 2\pi /f)$ with $\pi  =  T^a \pi^a$, $T^a  =  \lambda ^a /
2 (a=1,2,...8)$. External gauge fields are introduced via $v_\mu$
and $a_\mu$. The $\chi$ is defined by $\chi = 2B_0 (\cal S + \it i
\cal P )$. Explicit chiral symmetry breaking due to current quark
masses can be introduced by treating those masses as if they were
uniform external scalar field $S$\cite{Bi96}.

The non-linear realization of chiral symmetry is expressed in
terms of $u=\sqrt{U}$ and $h = h(u,g_R, g_L )$ defined from $ u
\rightarrow g_R u h^\dagger = hug_L ^\dagger $. In this
realization, we naturally have the following covariant quantities
\begin{eqnarray}
i\Gamma_\mu &=& \frac{i}{2} ( u^\dagger \partial_\mu u + u
\partial_\mu u^\dagger ) +\frac{1}{2} u^\dagger (v_\mu + a_\mu )u
+ \frac{1}{2}u(v_\mu - a_\mu )u^\dagger ~, \nonumber \\ i
\Delta_\mu &=& \frac{i}{2} ( u^\dagger \partial_\mu u - u
\partial_\mu u^\dagger ) +\frac{1}{2} u^\dagger ( v_\mu + a_\mu )u
- \frac{1}{2}u( v_\mu - a_\mu )u^\dagger  ~, \nonumber \\ \chi_+
&=& u^\dagger \chi u^\dagger + u \chi u~,
\end{eqnarray}
whose transformations are carried out in terms of $h$, i.e.,
$\Gamma_\mu \rightarrow h \Gamma_\mu h^\dagger -\partial_\mu h
\cdot h^\dagger $, $\Delta_\mu \rightarrow h \Delta_\mu h^\dagger
$, and $\chi_+ \rightarrow h\chi_+ h^\dagger$. With these
quantities, the Lagrangian in eq.(2) can be rewritten as
\begin{equation}
{\cal L} ( \pi )= f^2 \langle i\Delta_\mu i\Delta^\mu \rangle
+\frac{f^2}{4} \langle \chi_+ \rangle .
\end{equation}
As for the massive spin-1 mesons, we include only the mass and
kinetic terms\cite{Pa96}
\begin{equation}
{\cal L} (V, A) =m_V ^2 \langle (V_\mu - {{i \Gamma_\mu}\over g}
)^2 \rangle + m_A ^2 \langle (A_\mu - {{i r \Delta_\mu}\over g}
)^2 \rangle - \frac{1}{2} \langle (^G V_{\mu\nu})^2 \rangle
-\frac{1}{2} \langle {( A_{\mu\nu}) }^2 \rangle
\end{equation}
with
\begin{eqnarray}
^G V_{\mu\nu} &=& \partial_\mu V_\nu - \partial_\nu V_\mu - ig[
V_\mu, V_\nu]-iG[A_\mu, A_\nu]~, \nonumber \\ A_{\mu\nu} &=&
\partial_\mu A_\nu - \partial_\nu A_\mu - ig[ V_\mu,
A_\nu]-ig[A_\mu, V_\nu]\ ~,
\end{eqnarray}
where $V_\mu = T^a V_\mu ^a $($A_\mu = T^a A_\mu ^a $) denotes
spin-1 vector (axial-vector) meson field and $g$ denotes a $V \pi
\pi$ coupling constant. The chiral transformation rules of spin-1
fields are expressed in terms of $h$
\begin{equation}
V_\mu \rightarrow hV_\mu h^\dagger - \frac{i}{g}\partial_\mu h
\cdot h^\dagger ~,~ A_\mu \rightarrow h A_\mu h^\dagger.
\end{equation}
Note that we have introduced a new form of $^G V_{\mu\nu}$. The
chiral symmetry is preserved for any value of $G$ at the chiral
limit in $^G V_{\mu\nu}$, so that the value of $G$ cannot be
determined from the chiral symmetry. If $G$ is equal to $g$ as in
the HGS approach\cite{bando}, the result may reproduce
experimental data by explicitly including other higher order
terms.

The introduction of the $\cal L_{S}$ term can be found in
ref.\cite{Sh00}. The resulting Lgarangian is given as
\begin{equation}
{\cal L}_S \sim  - \frac{1}{2} (\frac{s_m}{f})^2 (\tilde M)_a ^2
(\pi^a)^2 + \frac{1}{2}s_m M_a j^{a} ,
\end{equation}
where $\tilde M_a^2 = \frac{1}{6}(2B_0\alpha)^2\delta_{8a} +
M_a^2$.

For the mixing between axial vector mesons and pion fields, we
define $A_{\mu}^{'}$ as
\begin{eqnarray}
A_\mu^{a} &=&  A^{\prime a}_\mu - \frac{r}{gf} \partial_\mu
\pi^{a} +\frac{r}{gf}f_{abc} \pi^{b}B_{\mu}^{c}\nonumber\\ &=&
A^{\prime a}_\mu +i\frac{r}{g}\Delta_{\mu}^{a},
\end{eqnarray}
where $B_{\mu}$ denotes a photon field and $Q=T^3 + \frac{Y}{2}$.
As for the mixing between vector mesons and pion fields, we also
define $V_{\mu}^{'}$ as
\begin{equation}
V_\mu^{a} = V^{\prime a}_\mu -
\frac{Gr^2}{2g^2f^2}f_{abc}\pi^{b}\partial_\mu \pi^{c} .
\end{equation}

Besides the above mixings, we introduced the $\rho-\omega$ mixing
and the mixing between the vector meson and photon field. Since
both mixings are explained in detail in ref.\cite{Sh00}, we skip
them here.

This field redefinition of eq.(10), which differs from our
previous one, leads to an explicitly manifest gauge invariance of
the amplitude relevant to the radiative decay of the axial vector
meson. But it does not give any differences in $V-\pi-\gamma$
reaction in our previous investigations, and describes correctly $
A-V-\pi$ and $ A-\gamma-\pi$ reactions. The lagrangian is, then,
simply summerized as
\begin{eqnarray}
{\cal
L}&=&\frac{1}{2}m_{Va}^2V_{\mu}V^{\mu}+\frac{1}{2}m_{Aa}^2A_{\mu}A^{\mu}\nonumber\\
&
&+\frac{m_{Va}^2}{2gf_{a}^2}(1-\frac{Gr^2}{g})f_{abc}V_{\mu}^{a}\pi^{b}\partial^{\mu}\pi^{c}\nonumber\\
& &+eQf_{abc}B_{\mu}^{a}\pi^{b}\partial^{\mu}\pi^{c}\nonumber\\ &
&-\frac{1}{4}(\partial_{\mu}V_{\nu}-\partial_{\nu}V_{\mu})^2
-\frac{1}{4}(\frac{e}{g})^{2}Q^{2}(\partial_{\mu}B_{\nu}-\partial_{\nu}B_{\mu})^2\nonumber\\
&
&-\frac{e}{2g}Q(\partial_{\mu}V_{\nu}-\partial_{\nu}V_{\mu})(\partial^{\mu}B^{\nu}-\partial^{\nu}B^{\mu})\nonumber\\
&
&+\frac{1}{2}\frac{Gr}{gf}f_{abc}(\partial_{\mu}V_{\nu}^{a}-\partial_{\nu}V_{\mu}^{a})
(A^{\mu b}\partial^{\nu}\pi^{c}+\partial^{\mu}\pi^{b}A^{\nu
c})\nonumber\\ &
&+\frac{1}{2}\frac{Gr}{gf}(\frac{e}{g})Qf_{abc}(\partial_{\mu}B_{\nu}^{a}-\partial_{\nu}B_{\mu}^{a})
(A^{\mu b}\partial^{\nu}\pi^{c}+\partial^{\mu}\pi^{b}A^{\nu
c})\nonumber\\ &
&-\frac{1}{4}(\partial_{\mu}A_{\nu}-\partial_{\nu}A_{\mu})^2\nonumber\\
&
&+\frac{1}{2}\frac{r}{f}f_{abc}(\partial_{\mu}A_{\nu}^{a}-\partial_{\nu}A_{\mu}^{a})
(V^{\mu b}\partial^{\nu}\pi^{c}+\partial^{\mu}\pi^{b}V^{\nu
c})\nonumber\\ &
&-\frac{1}{2}\frac{r}{f}\frac{e}{g}Qf_{abc}(\partial_{\mu}A_{\nu}^{a}-\partial_{\nu}A_{\mu}^{a})
\pi^{b}(\partial^{\mu}B^{\nu c}-\partial^{\nu}B^{\mu
c})\nonumber\\ & &-\frac{1}{2}m_{\pi
a}^2\pi^{a}\pi^{a}+\frac{1}{2}\partial_{\mu}\pi^{a}\partial^{\mu}\pi^{a},
\end{eqnarray}
where $m_{Va}^2=g^2(f_V^2 + s_ms_vM_a)$, $m_{Aa}^2=g^2(f_A^2 +
s_ms_a M_a)$and $V_{\mu}$ and $A_{\mu}$ stand for redefined fields
$V_{\mu}^{'}$ and ${A_{\mu}}^{'}$. The relation of our chiral
effective lagrangian to the other lagrangians was discussed at the
ref.\cite{Sh00, Bi96}. The 9 th and 12 th terms, and the 8 th and
11 th terms in the above lagrangian, which were omitted in our
previous lagrangian, corresponds to the $ A-\gamma-\pi$ and $
A-V-\pi$ interactions, respectively. Detailed discussions
concerning the lagranigans and applications to decay modes will be
done at the next section. In order to determine pseudoscalar meson
mass and decay constants, we exploit the following covariant
quantities \ba m_{\pi
a}^2&=&(M_{a}+(\frac{s_m}{f})^2\tilde{M_{a}}^2)/Z_{\pi a}^2~,~ f_a
= Z_{\pi a}f \nonumber\\ &with~~&  Z_{\pi a}^2 = (1+s_m
s_d\frac{M_a}{f^2})~. \ea

Mass splitting between non-strange particles and strange particles
is generated from the interaction of the meson fields with scalar
particle fields.

\section{Axial vector meson decay}

\idn The Lagrangian for $ A-V-\pi$ reaction is given as
\begin{eqnarray}
{\cal
L}_{AV\pi}&=&\frac{1}{2}\frac{Gr}{gf}f_{abc}(\partial_{\mu}V_{\nu}^{a}-\partial_{\nu}V_{\mu}^{a})
(A^{\mu b}\partial^{\nu}\pi^{c}+\partial^{\mu}\pi^{b}A^{\nu
c})\nonumber\\
&+&\frac{1}{2}\frac{r}{f}f_{abc}(\partial_{\mu}A_{\nu}^{a}-\partial_{\nu}A_{\mu}^{a})
(V^{\mu b}\partial^{\nu}\pi^{c}+\partial^{\mu}\pi^{b}V^{\nu c}).
\end{eqnarray}

The 1st term corresponds to the lagrangian used by
Xiong\cite{Br92} and Haglin\cite{Ha94}. The 2nd term comes from
the $\partial_{\mu}\pi$ in the mixing of eq.(10), which avoids a
direct coupling of the pion to the axial vector mesons. From the
above Lagrangian, the partial decay width of $a_{1}$ meson to
$\rho\pi$ is calculated as follows
\begin{eqnarray}
\Gamma_{a_{1}^{\pm}\rightarrow\rho\pi}&=&\Gamma_{a_{1}^{\pm}\rightarrow\rho^{\pm}\pi^{0}}+\Gamma_{a_{1}^{\pm}\rightarrow\rho^{0}\pi^{\pm}}\nonumber\\
\Gamma_{a_{1}^{\pm}\rightarrow\rho^{\pm}\pi^{0}({a_{1}^{\pm}\rightarrow\rho^{0}\pi^{\pm}})}&=&\frac{1}{24\pi}\frac{|\vec{q}~|}{m_{a}^2}|{\cal
M}_{a_{1}^{\pm}\rightarrow\rho^{\pm}\pi^{0}({a_{1}^{\pm}\rightarrow\rho^{0}\pi^{\pm}})}|^2~,
\end{eqnarray}
where $\vec{q}$ is a incoming pion momentum in the $a_{1}$ rest
frame given as ${\vec{q}}~^2=
\frac{(m_{a_{1}}^2+m_{\pi}^2-m_{\rho}^2)^2}{4m_{a_{1}}^2}-m_{\pi}^2$.
Invariant amplitude ${\cal M}[{a_{1}(p,
~\epsilon_{A})\rightarrow\rho(k,~\epsilon_{V}) \pi(q)}]$ and
$|{\cal M}|^2$ are given as
\begin{eqnarray}
{\cal M}_{a_{1}\rightarrow \rho \pi}=f_{1} \epsilon_{V\nu}((q\cdot
k)g^{\nu}_{\mu}-k_{\mu}q^{\nu})\epsilon_{A}^{\mu}+f_{2}\epsilon_{V\nu}((p\cdot
q)g^{\nu}_{\mu}-p_{\mu}q^{\nu})\epsilon_{A}^{\mu}~,
\end{eqnarray}
\begin{eqnarray}
|{\cal M}_{a_{1}^{\pm}\rightarrow\rho^{0}\pi^{\pm}}|^2=|{\cal
M}_{a_{1}^{\pm}\rightarrow\rho^{\pm}\pi^{0}}|^2&=&4[f_{1}(2(k\cdot
q)^2+m_{\rho}^2(m_{\pi}^2+{\vec{q}}~^2))\nonumber\\& &
+f_{2}(2(p\cdot q)^2+(q\cdot k)^2)+6f_{1}f_{2}(p\cdot q)(k\cdot
q)].\nonumber\\
\end{eqnarray}
Coupling constants $f_{1}, f_{2}$ are defined as
$f_{1}=\frac{1}{2}\frac{Gr}{gf}$, $f_{2}=\frac{1}{2}\frac{r}{f}$.
$p,~q$ and $k$ stand for the momenta of the axial-vector meson,
the vector meson, and the pion, respectively. The partial decay
width of $K_{1}$ meson to $\rho K$ is also calculated in the same
manner,
\begin{eqnarray}
\Gamma_{K_{1}^{\pm}\rightarrow\rho
K}&=&\Gamma_{K_{1}^{\pm}\rightarrow\rho^{\pm}
K^{0}}+\Gamma_{K_{1}^{\pm}\rightarrow\rho^{0} K^{\pm}}\nonumber\\
\Gamma_{K_{1}^{\pm}\rightarrow\rho
K}&=&\frac{1}{24\pi}\frac{|~\vec{q}~|}{m_{K_1}^2}|{\cal
M}_{K_{1}^{\pm}\rightarrow\rho K}|^2~,
\end{eqnarray}
where invariant amplitude ${\cal M}[{K_{1}(p,
~\epsilon_{A})\rightarrow\rho(k,~\epsilon_{V}) K(q)}]$ and $|{\cal
M}|^2$ are given as
\begin{eqnarray}
{\cal M}_{K_{1}\rightarrow \rho K}=f_{K_{1}}
\epsilon_{V\nu}((q\cdot
k)g^{\nu}_{\mu}-k_{\mu}q^{\nu})\epsilon_{A}^{\mu}+f_{K_{2}}\epsilon_{V\nu}((p\cdot
q)g^{\nu}_{\mu}-p_{\mu}q^{\nu})\epsilon_{A}^{\mu}~,
\end{eqnarray}

\begin{eqnarray}
|{\cal M}_{K_{1}^{\pm}\rightarrow\rho^{0}
K^{\pm}}|^{2}&=&[f_{K_{1}}(2(k\cdot
q)^2+m_{\rho}^2(m_{K^{\pm}}^2+{\vec{q}}~^2))\nonumber\\&
&+f_{K{2}}(2(p\cdot q)^2+(q\cdot k)^2)+6f_{K_{1}}f_{K{2}}(p\cdot
q)(k\cdot q)]\nonumber\\ |{\cal
M}_{K_{1}^{\pm}\rightarrow\rho^{\pm}
K^{0}}|^{2}&=&2[f_{K_{1}}(2(k\cdot
q)^2+m_{\rho}^2(m_{K^{0}}^2+{\vec{q}}~^2))\nonumber\\&
&+f_{K{2}}(2(p\cdot q)^2+(q\cdot k)^2)+6f_{K_{1}}f_{K{2}}(p\cdot
q)(k\cdot q)] ~,
\end{eqnarray}
where $f_{K_{1}},f_{K_{2}}$ are defined as
$f_{K_{1}}=\frac{1}{2}\frac{Gr}{gf_K}$,
$f_{K_{2}}=\frac{1}{2}\frac{r}{f_{K}}$, with kaon decay constant
$f_K$. In this case, incoming momentum $\vec{q}$ is
$\vec{q}~^2=\frac{(m_{K_{1}}^2+m_{K}^2-m_{\rho}^2)^2}{4m_{K_{1}}^2}-m_{K}^2$,
and $p,~q,~k$ are $K_{1}$ meson, $\rho$ meson, kaon momentum,
respectively.


On the other hand, the lagrangian for $ A-\gamma-\pi$ reaction is
obtained from eq. (12) as follows
\begin{eqnarray}
{\cal
L}_{A\gamma\pi}&=&\frac{1}{2}\frac{Gr}{gf}(\frac{e}{g})Qf_{abc}(\partial_{\mu}B_{\nu}^{a}-\partial_{\nu}B_{\mu}^{a})
(A^{\mu b}\partial^{\nu}\pi^{c}+\partial^{\mu}\pi^{b}A^{\nu
c})\nonumber\\ &-&
\frac{1}{2}\frac{r}{f}(\frac{e}{g})Qf_{abc}(\partial_{\mu}A_{\nu}^{a}-\partial_{\nu}A_{\mu}^{a})
\pi^{b}(\partial^{\nu}B^{c}-\partial^{\mu}B^{c})~.
\end{eqnarray}

The 1st term in lagrangian of eq.(21) resembles the lagrangian of
Xiong\cite{Br92}, but in our lagrangian, the 2nd term appears,
which originated from the 3rd term in the mixing of eq.(10). This
mixing term allows a direct coupling of the photon to the pion and
causes small deviation from the conventional vector meson
dominance (VMD) model (see the Figure 2 and 3 in ref.\cite{Sh00}).
Moreover it makes the relevant amplitude gauge invariant as shown
below.

Partial decay width of $a_{1}$ meson to $\gamma \pi$ and $K_{1}$
meson to $\gamma K$ are expressed as
\begin{eqnarray}
\Gamma_{a_{1}^{\pm} \rightarrow \gamma
\pi^{\pm}}&=&\frac{1}{24\pi}\frac{|\vec{q}|}{m_{a_{1}}^2}|{\cal
M}_{a_{1}^{\pm}\rightarrow\gamma \pi^{\pm}}|^2 \nonumber\\
\Gamma_{K_{1}^{\pm} \rightarrow \gamma
K^{\pm}}&=&\frac{1}{24\pi}\frac{|\vec{q}|}{m_{K_{1}}^2}|{\cal
M}_{K_{1}^{\pm}\rightarrow\gamma K^{\pm}}|^2 ~,
\end{eqnarray}
where invariant amplitude ${{\cal M}[{a_{1}(p,
~\epsilon_{A})\rightarrow\gamma(k,~\epsilon_{\gamma}) \pi(q)}]}$
and $|{\cal M}|^2$ are given as
\begin{eqnarray}
{\cal M}_{a_{1}\rightarrow \pi \gamma}=h_{1}
\epsilon_{\gamma\nu}((q\cdot
k)g^{\nu}_{\mu}-k_{\mu}q^{\nu})\epsilon_{A}^{\mu}-h_{2}\epsilon_{\gamma\nu}((p\cdot
k)g^{\nu}_{\mu}-k_{\mu}p^{\nu})\epsilon_{A}^{\mu}~,
\end{eqnarray}

\begin{eqnarray}
|{\cal M}_{a_{1}^{\pm} \rightarrow \gamma \pi^{\pm}(K_{1}^{\pm}
\rightarrow \gamma K^{\pm}) }|^{2}=4[2h_{1}(k\cdot
q)^2+2h_{2}(k\cdot p)^2-4h_{1}h_{2}(k\cdot q)(k\cdot p)]~,
\end{eqnarray}
where $\vec{q}~^2=(\frac{m_{a_1}^2-m_{\pi}^3}{2m_{a_1}})^2$ and
$h_{1}=\frac{e}{g}f_{1},h_{2}=\frac{e}{g}f_{2}$. $p,~q$ and $k$
are axial-vector meson, vector meson, photon momentum,
respectively. It is easy to check that the above amplitude is
fully gauge invariant.

By exploiting the above equations, we determine values of $G$, $r$
and $g$ by fitting our numerical predictions of eqs. (15), (18)
and (22) to the experimental data.
The results and experimental data are tabulated in table 1 with
the comparison to other cases. The masses of axial vector meson
used here are $a_{1}(1260)=~1210 \rm{MeV}$ , $ K_{1}(1270)=~1280
\rm{MeV}$, respectively.

As shown in table 1, the theoretical value of
$\Gamma_{a_{1}\rightarrow\pi\gamma}$ in Haglin's paper\cite{Ha94}
is overestimated about 2 times. Consequently,
$\Gamma_{K_{1}\rightarrow K\gamma}$ in our prediction is much
smaller than that of Haglin. It leads to a smaller contribution of
$K_{1}$ to the photon spectrum compared to that of $a_{1}$ as will
be shown.

\section{Photon Production Rate}

\idn In this section, we show a brief formalism regarding the
emitted photon spectrum in a hadronic gas.
In principle, including the axial vector mesons as intermediate
states can add other channel processes in photon production.
However, we consider only the s-channel contribution as in figure
1, because the contributions of other channels turned out to be
small compared to that of s-channel in the hadronic gas of
$T=100\sim200 \rm{MeV}$\cite{Br92,ko96}.

For a reaction of the mesons, $1\,+\,2\,\rightarrow\,3\,+\gamma$
the photon production rate with temperature $T$ \cite{Sa97} is
given by
\begin{eqnarray}
E\frac{dR}{d^3p}&=&\frac{{\cal{N}}}{16(2\pi)^7E}\,\int_{(m_1+m_2)^2}^{\infty}
\,ds\,\int_{t_{{\s min}}}^{t_{{\s max}}}\,dt\,|{\cal M}|^2\,
\int\,dE_1\nonumber\\
&&\times\int\,dE_2\frac{f(E_1)\,f(E_2)\left[1+f(E_3)\right]}{\sqrt{aE_2^2+
2bE_2+c}},
\end{eqnarray}
where
\begin{eqnarray}
a&=&-(s+t-m_2^2-m_3^2)^2~,\nonumber\\
b&=&E_1(s+t-m_2^2-m_3^2)(m_2^2-t)+E_{\gamma}[(s+t-m_2^2-m_3^2)(s-m_1^2-m_2^2)\nonumber\\
&&-2m_1^2(m_2^2-t)]~,\nonumber\\
c&=&-E_1^2(m_2^2-t)^2-2E_1E_{\gamma}[2m_2^2(s+t-m_2^2-m_3^2)-(m_2^2-t)(s-m_1^2-m_2^2)]
\nonumber\\
&&-E_{\gamma}^2[(s-m_1^2-m_2^2)^2-4m_1^2m_2^2]-(s+t-m_2^2-m_3^2)(m_2^2-t)\nonumber\\
&&\times(s-m_1^2-m_2^2)
+m_2^2(s+t-m_2^2-m_3^2)^2+m_1^2(m_2^2-t)^2~,\nonumber\\ E_{1{{\s
min}}}&=&\frac{(s+t-m_2^2-m_3^2)}{4E_{\gamma}}+
\frac{E_{\gamma}m_1^2}{s+t-m_2^2-m_3^2}~,\nonumber\\ E_{2{{\s
min}}}&=&\frac{E_{\gamma}m_2^2}{m_2^2-t}+\frac{m_2^2-t}{4E_{\gamma}}~,\nonumber\\
E_{2{{\s max}}}&=&-\frac{b}{a}+\frac{\sqrt{b^2-ac}}{a}~,\nonumber
\end{eqnarray}
where ${\cal N}$ is the overall degeneracy of the particles 1 and
2, ${\cal M}$ is the invariant amplitude of the reactions
considered in this paper (summed over final states and averaged
over initial states), and $s$, $t$, $u$ are the usual Mandelstam
variables. In the above equation, the indices 1, 2, 3 and $\gamma$
mean for the incident pion, incident $\rho$ meson, outgoing pion
and outgoing photon, respectively. They are allowed to vary in a
whole range to take off-shell property of the relevant particles
into account. $f(E)=\frac{1}{(e^{E/T}-1)}$ is the Bose-Einstein
distribution function.

The invariant amplitude ${\cal M}$ for the diagram in figure 1 is
calculated as
\begin{eqnarray} {\cal M}&=&2\epsilon_{V
\alpha}[f_{1}((q\cdot
k)g^{\alpha}_{\mu}-k_{\mu}q^{\alpha})+f_{2}((p\cdot
q)g^{\alpha}_{\mu}-p_{\mu}q^{\alpha})]D^{\mu \beta}\nonumber\\ &
&\times 2 \epsilon_{\gamma \nu}[h_{1}((i\cdot
j)g^{\nu}_{\beta}-j_{\beta}i^{\nu})-h_{2}((p\cdot
j)g^{\nu}_{\beta}-j_{\beta}p^{\nu })]~,
\end{eqnarray}
where $k$ and $q$ are momenta of the incoming $\rho$ and $\pi$
mesons, $i, j$ are outgoing photon and $\pi$ meson momentum and
$p$ is the axial vector meson's momentum, respectively.
$D^{\mu\beta}$ is the propagator for the axial vector meson
\begin{equation}
D^{\mu\beta}=(g^{\mu\beta}-p^{\mu}p^{\beta})\frac{1}{p^2-m_{a_1(K_{1})}^2-I
m_{a_{1}(K_{1})}\Gamma_{a_{1}(K_{1})}}~.
\end{equation}

The processes we are going to study are  $\pi\rho\rightarrow
a_{1}\rightarrow \pi\gamma$ and $K\rho\rightarrow K_{1}\rightarrow
K\gamma$ in figure 1. Using the approximation by W. Greiner \cite
{Gr86}, we numerically compute a three dimensional integral in
eq.(25) and get the photon spectra at various temperatures.
Results are shown in figure 2. $a_{1}$ and  $K_{1}$ resonance's
contributions are presented as the solid and dot-dashed lines,
respectively. The spectra are calculated at three different
temperatures, $T=200, 150, 100~ \rm MeV$ from the uppermost,
respectively. To investigate a dependence of the spectrum on the
given lagrangian, previous results \cite{ Br92,Ha94}, which are
reproduced by switching off the 2nd terms in eqs.(14) and (21),
are presentd by the dotted lines. Likewise to the discussion in
reference\cite{ko96}, there does not appear such a discernable
difference due to the different lagrangian. Namely, in the case of
$a_{1}$, our results show predictions more or less similar to
those of Haglin. But maximum values in the spectrum are increased
within 10$\%$ at each temperature. It is also noticeable that the
peak positions of $E_{\gamma}$ in the spectrum are shifted a
little bit backward compared to the previous predictions.

The $K_{1}$ contribution, as expected, turned out to be smaller in
high $E_{\gamma}$ region at least one order than that of $a_{1}$.
But at low $E_{\gamma}$ region, it shows a comparable behavior to
that of $ a_{1}$. Therefore, in the low $E_{\gamma}$ region below
0.5 GeV, both $a_{1}$ and $K_{1}$ axial vector mesons show
competitive roles in this spectrum. But the $K_{1}$ meson's role
in the high $E_{\gamma}$ region is decreased up to $3\sim 4$ GeV
region.

\section{Conclusion}
\idn Based on our previous $SU_{L} (3)\otimes SU_{R} (3)$ chiral
lagrangian\cite{Sh00}, $K_{1}$ and $a_{1}$ meson's contributions
to $K\rho \rightarrow K_{1} \rightarrow K\gamma$ and $\pi\rho
\rightarrow a_{1} \rightarrow \pi\gamma$ channels are investigated
for the emitted photon spectrum in a hot hadronic matter. Before
the calculation, the relevant decay widths are quite well
reproduced within experimental errors. In specific, the radiative
and hadronic decays of the axial vector meson $a_{1}$ and $K_{1}$
are shown to be consistently explained in our lagrangian. Our
emitted phton spectrum shows that $K_{1}$ and $a_{1}$ mesons could
be dominant channels in low $E_{\gamma}$ region below 0.5 GeV,
while the role of $K_{1}$ meson is decreased in high $E_{\gamma}$
region.

\newpage


\begin{table}
\begin{center}
\setlength{\tabcolsep}{0mm}
\begin{tabular}{|c|c|c|c|c|c|}\hline
  &~Xiong$^{\cite{Br92}}$, Haglin$^{\cite{Ha94}}$~ & ~C. Song$^{\cite{So93}}$ ~&~ P. Ko $^{\cite{ko96}}$~&~Ours~ & Experiment \\\hline
  $m_{a_{1}}(1260)$ & 1230 & 1230 & 1260 & 1210 & ~1230 $\pm$ 40 $\rm{MeV}$~ \\

  $m_{K_{1}}(1270)$ & 1273 & $\cdot$  & $\cdot$ & 1280 & 1273$\pm$7 $\rm{MeV}$
  \\\hline
  $\Gamma_{a_{1}\rightarrow \rho\pi}$ & 400 & 400 & 328 & 488 & 200$\sim$600$\rm{MeV}$ \\
  $\Gamma_{a_{1}\rightarrow \pi\gamma}$ & 1.4 & $\cdot$ & 0.67 & ~0.688~ & 0.64$\pm$0.28$\rm{MeV}$ \\
  $\Gamma_{K_{1}\rightarrow \rho K}$ & 37.8 &$\cdot$  &$\cdot$  & 47  & 57$\pm$5 $\rm{MeV}$  \\
  $\Gamma_{K_{1}\rightarrow K\gamma}$ & 1.5 & $\cdot$ &$\cdot$  & 0.350 &  \\ \hline
\end{tabular}
\end{center}
\vspace{5mm} \caption[Table 1]{Masses and decay widths of the
relevant axial vector mesons.} \label{tab1}
\end{table}
\vspace{5mm}

\newpage

\begin{figure}
\begin{center}
\epsfysize5cm \leavevmode\epsfbox{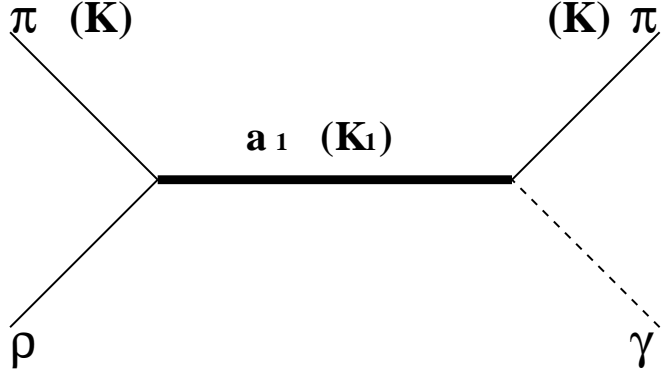} \caption{Feynman
diagram of $\pi\rho \rightarrow \pi\gamma$($K\rho \rightarrow
K\gamma$) through $a_{1}(K_{1})$ resonance}
\end{center}
\end{figure}

\begin{figure}
\begin{center}
\epsfysize10cm \leavevmode\epsfbox{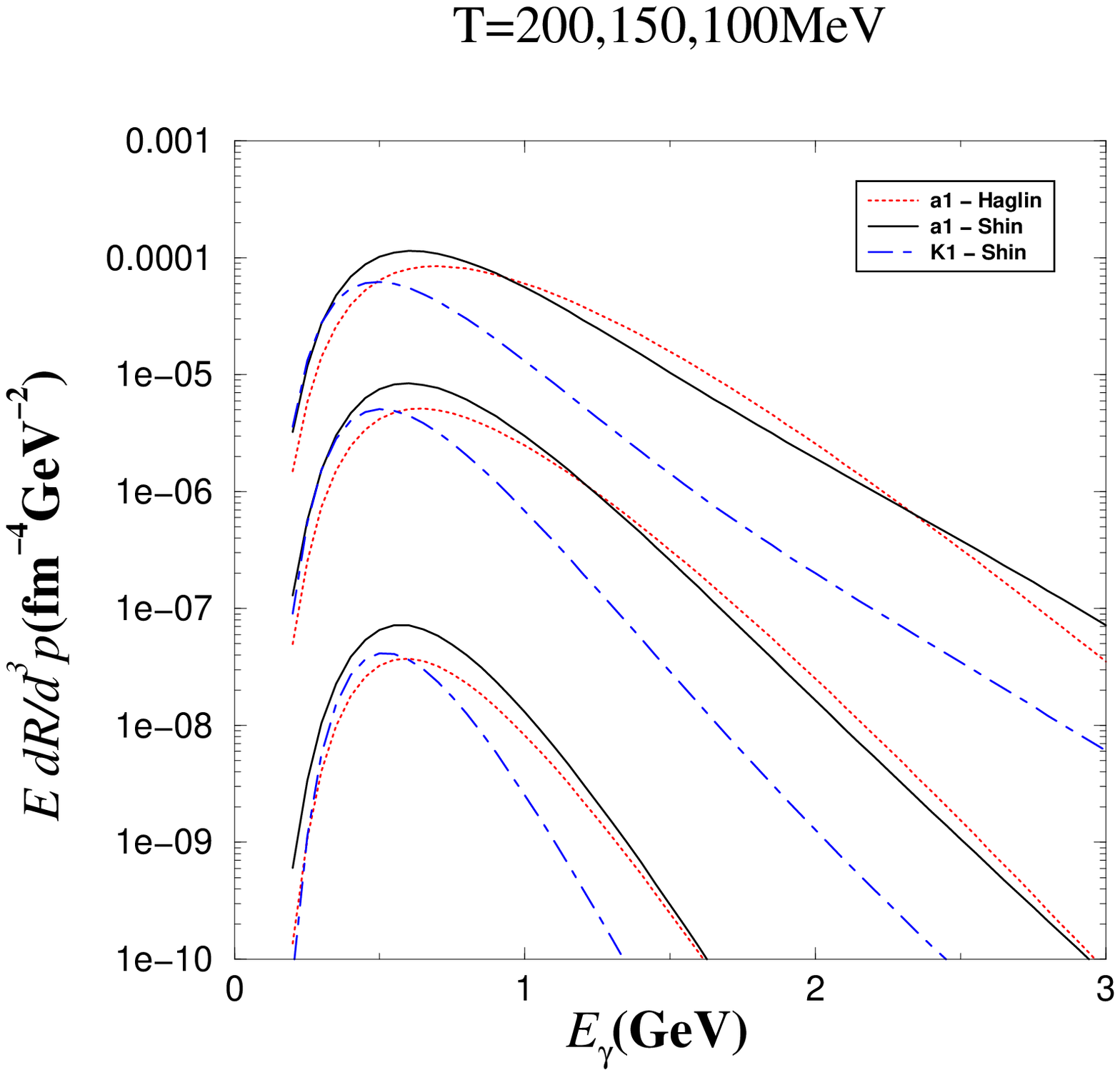} \caption{Photon
production rate at T=100 $\sim$ 200 MeV. From the uppermost
$T=200,~150,~100$, respectively.}
\end{center}
\end{figure}

\end{document}